\documentclass[prb,aps,twocolumn,amsmath,amssymb,floatfix,
superscriptaddress]{revtex4}

\usepackage[dvips]{graphics}
\usepackage{color}
\definecolor{dred}{rgb}{0,0,0.6}

\begin{document}

\title{Logical operations using phenyl ring}

\author{Moumita Patra}

\affiliation{Physics and Applied Mathematics Unit, Indian Statistical
Institute, 203 Barrackpore Trunk Road, Kolkata-700 108, India}

\author{Santanu K. Maiti}

\email{santanu.maiti@isical.ac.in}

\affiliation{Physics and Applied Mathematics Unit, Indian Statistical
Institute, 203 Barrackpore Trunk Road, Kolkata-700 108, India}

\begin{abstract}

Exploiting the effects of quantum interference we put forward an idea of 
designing three primary logic gates, OR, AND and NOT, using a benzene
molecule. Under a specific molecule-lead interface geometry, anti-resonant 
states appear which play the crucial role for AND and NOT operations, while 
for OR gate no such states are required. Our analysis leads to a possibility 
of designing logic gates using simple molecular structure which might be 
significant in the area of molecular electronics.

\end{abstract}

\maketitle

Organic molecules are considered as the basic building blocks of designing
nano-electronic devices~\cite{r1,r2,r3,r4,r5,r6,r7,r8,r9}. Especially, the 
molecules having single or multiple loop geometries are the most promising 
candidates due to the fact that the effects of quantum interference can be 
directly implemented 
in such systems~\cite{r10,r11,r12,r13,r14,r15,ojeda1,ojeda2,ojeda3}. 
Following the idea of Aviram 
and Ratner~\cite{r17}, interest in the subject of electron transport through 
molecular structures has rapidly picked up with considerable theoretical and 
experimental works~\cite{r1,r2,r3,r4,r5,r6,r7,r8,r9,r10,r11,r12,r13,r14,r15,
r18,r19,r20,r21,r22}. Several propositions have been made for designing 
different molecular based electronic devices like transistor~\cite{r23}, 
rectifier~\cite{r24}, switches~\cite{r25,r25a,r25b,r25c}, and to name a few. 
Comparatively much less effort has been given in designing molecular logic 
gates~\cite{r26,r27,r28,r29}, particularly, considering simple molecular 
structures. Here it is important to note that though one of the authors of 
us has previously suggested the possibilities of designing logic gates 
using quantum rings but in all those cases large magnetic flux is 
required~\cite{new1,new2,new3,new4} and confining a large magnetic flux 
in a small sized ring has always been a challenging task which we 
definitely want to avoid if possible.

Motivated by these facts, in the present work, we propose an idea of 
constructing all three basic logic gates (OR, AND and NOT) considering a 
benzene molecule exploiting the effects of quantum interference without
considering any magnetic flux. For OR and 
AND gates we use a three-terminal molecular junction, and for NOT gate a 
two-terminal setup is used. The main idea is that under a particular 
molecule-to-lead configuration the molecular junction exhibits 
{\em anti-resonant states}~\cite{ars1,ars2,r16} at some typical energies. 
These anti-resonant states, specific to interferometric geometry, play the 
key role behind 
AND and NOT operations. For the OR gate operation no such states are required,
and thus, can easily be designed. We measure the response in terms of 
transmission probability. Finite (high) transmission in the output lead
represents `ON' state, while `OFF' state means zero (vanishingly small)
transmission.

To calculate transmission probability we use Green's function formalism 
where the effects of input and output leads are incorporated through 
self-energy correction~\cite{r9}. In terms of self-energies the effective 
Green's function of the molecule becomes~\cite{r9} 
$G^r=\left(E-H_M-\sum_{\mbox {\small p}} \Sigma_{\mbox {\small p}}
\right)^{-1}$, where $E$ is the energy of the injecting electron, $H_M$ 
is the molecular Hamiltonian and $\Sigma_{\mbox {\small p}}$ is the 
self-energy term due to lead-p (p can run from $1$ to $2$ or $1$ to $3$ 
depending on two- or three-terminal bridge set-up). The Hamiltonian $H_M$ is 
described within a tight-binding (TB) framework and it looks like~\cite{r14,
ars1,ars2,r16} 
$H_M=\sum_i \epsilon_i c_i^{\dagger} c_i + \sum_i t \left(c_{i+1}^{\dagger} 
c_i + h.c. \right)$, where $\epsilon_i$ and $t$ represent the on-site energy 
and nearest-neighbor hopping (NNH) integral, respectively, and $c_i^{\dagger}$ 
($c_i$) corresponds to the creation (annihilation) operator for an electron 
at $i$th site of the molecular ring.
Similar kind of TB Hamiltonians are also used to describe the side-attached
leads those are parameterized by on-site energy $\epsilon_0$ and NNH integral 
$t_0$. Once $G^r$ is formed, the two-terminal transmission probability between
lead-p and lead-q is determined from the relation~\cite{r9}
$T_{\mbox {\small pq}}=\mbox{Tr} \left[\Gamma_{\mbox {\small p}} 
G^r \Gamma_{\mbox {\small q}} G^a \right]$, 
where $G^a=(G^r)^{\dagger}$, and $\Gamma_{\mbox {\small p}}$'s are the
coupling matrices due to the coupling of the molecule with lead-p, where the
molecule-to-lead coupling is measured by $\tau_{\mbox {\small p}}$.

In what follows we present our results and discuss the logical operations. 
\begin{figure}[ht]
{\centering \resizebox*{7.75cm}{6.35cm}{\includegraphics{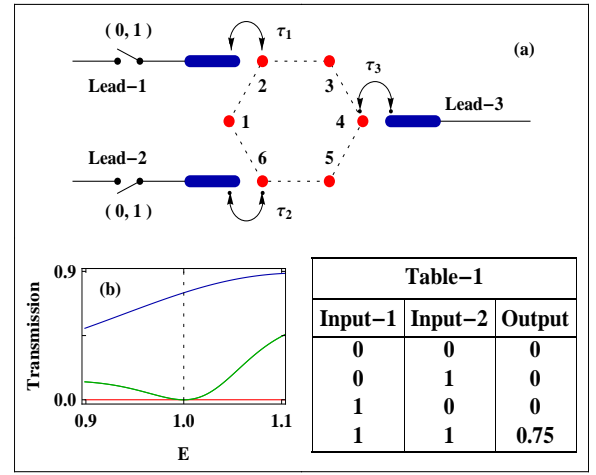}}\par}
\caption{(Color online). AND gate operation: (a) Schematic view of molecular
AND gate with two input (lead 1 and lead 2) and one output (lead 3) terminals. 
(b) Transmission probabilities as a function of energy for a small energy 
window, where the green and blue lines correspond to $T_{13}$ ($=T_{23}$) 
and $T_{13}+T_{23}$, respectively. The red line represents the case when 
both inputs are OFF. The AND gate response at $E=1\,$eV (marked by dashed 
line) is summarized in Table 1.} 
\label{fig1}
\end{figure}
The common set of parameter values which we fix for the numerical calculations
are as follows. In the molecule $\epsilon_i=0 \, \forall \, i$, unless 
otherwise stated, and $t=1\,$eV. In the leads $\epsilon_0=0$ and $t_0=2\,$eV, 
and the lead-to-molecule coupling $\tau_{\mbox {\small p}}$ is set at $1\,$eV.
Since we focus on the qualitative analysis rather than 
quantitative one, we choose this set of parameter values. All the physical 
phenomena remain unchanged for other choices of these parameters.

\vskip 0.1cm
\noindent
{\bf \underline{AND Gate}:} The AND gate operation is summarized in 
Fig.~\ref{fig1}. Figure~\ref{fig1}(a) represents the schematic diagram of
bridge setup, where each input terminal is connected via a toggle switch
which controls the `ON' and `OFF' states of the input signal. In the OFF
state electrons are no longer injected into the molecule from the input lead,
\begin{figure}[ht]
{\centering \resizebox*{7.75cm}{6.5cm}{\includegraphics{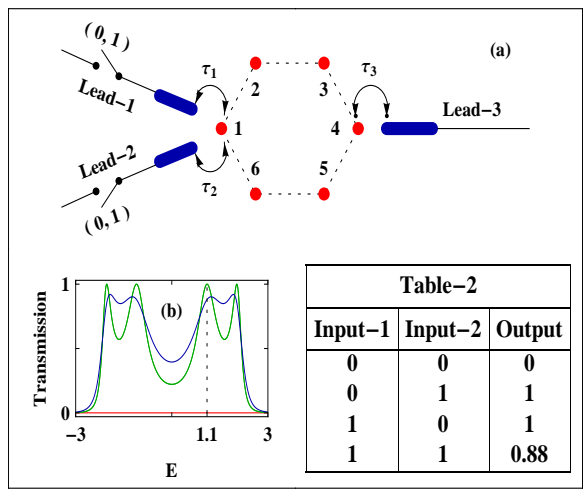}}\par}
\caption{(Color online). OR gate operation: (a) Molecular setup with two 
input and one output terminals for designing OR gate. (b) Transmission-energy
spectrum where different colored curves correspond to the identical meaning
as described in Fig.~\ref{fig1}(b). The OR gate response at $E=1.1\,$eV 
(marked by dashed line) is summarized in Table 2.} 
\label{fig2}
\end{figure}
and under this situation we assume that the lead is not connected to the 
molecular ring, and accordingly, we calculate effective Green's function 
$G^r$ considering the contributions from the rest leads. 

When both inputs are OFF naturally no response is obtained, as shown by the
red line of Fig.~\ref{fig1}(b). The situation becomes interesting when any 
one of the two inputs is ON. In the {\em meta connected} molecular junction 
(2-4 or 6-4 connection) anti-resonant states appear at $E=\pm 1\,$eV.
Here we concentrate at $E=+1\,$eV (we can also take
$E=-1\,$eV, as mentioned at the end of this section, since AND operation
is essentially obtained based on the energy location of anti-resonant
states), and therefore, we plot the
transmission-energy spectrum over a small energy window centering this
anti-resonant energy. From the green line of Fig.~\ref{fig1}(b) it is
clearly seen that $T_{13}$ ($=T_{23}$) provides a sharp dip (zero
transmission) at $E=1\,$eV, while it becomes finite at other energies.
This feature is associated with such an interferometric geometry~\cite{ars1,
ars2,r16,ars3} and not available in linear-like molecular structure.
When both the two inputs are ON, anti-resonant states disappear
and the net response (viz, $T_{13} + T_{23}$) becomes high (blue line).
Thus, if we think of an experimental setup where electrons having energy
$1\,$eV are injected or only one anti-resonant level having eigenenergy
$E=1\,$eV is placed within a narrow voltage window (that in principle can be
\begin{figure}[ht]
{\centering \resizebox*{8cm}{7.3cm}{\includegraphics{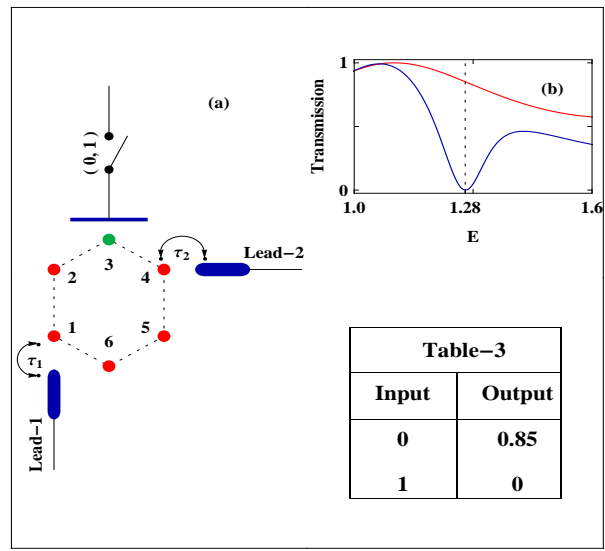}}\par}
\caption{(Color online). NOT gate operation: (a) Molecular setup for NOT 
gate where lead-1 and lead-2 are connected in {\em para configuration},
and they behave like source and drain electrodes as used in conventional
molecular junction. A gate electrode, which serves as input of NOT gate,
is placed in the vicinity of site $3$ through which its site potential 
($\epsilon_3$) can be altered. When the gate electrode is OFF, $\epsilon_3$ 
remains same (i.e., $\epsilon_3=0$) with other site energies of the phenyl 
ring, while it becomes finite under ON state. For ON state we set 
$\epsilon_3=1\,$eV. (b) Variation of $T_{12}$ as a function of energy 
where the red and blue lines correspond to $\epsilon_3=0$ and $1\,$eV,
respectively. The NOT gate operation at $E=1.28\,$eV (marked by dashed 
line) is summarized in Table 3.} 
\label{fig3}
\end{figure}
adjusted by suitable gate electrodes) then a clear AND gate operation is 
expected at this typical energy (i.e., $E=1\,$eV). This is exactly shown 
in Table 1. An identical AND gate response will also be
observed by setting the typical energy $E=-1\,$eV, since transmission-energy
spectrum becomes symmetric around the ring site energy (call it as 
$\epsilon$ since $\epsilon_i$'s are same for all sites of the molecular ring) 
which is fixed at zero. For identical molecule-to-lead coupling, the 
anti-resonant states appear at $E=\epsilon \pm t$~\cite{ars1}. Therefore, 
change of 
site energy or NNH integral or both simply shifts the whole transmission 
spectrum along with anti-resonant states, but the AND gate response at these
anti-resonant energies remains exactly invariant. Thus, no new physical 
phenomenon is expected by changing the parameter values.

\vskip 0.1cm
\noindent
{\bf \underline{OR Gate}:} Designing of an OR gate is rather quite simple
than any other gate as it does not involve any such anti-resonant states like
above. Only thing is that we have to set a typical energy where resonant 
transmission is obtained when any one or both inputs are high, and definitely
it will be the best if we can achieve maximum transmission probability close
to unity. The schematic representation of the molecular OR gate is shown in 
Fig.~\ref{fig2}(a) where two inputs are coupled to site $1$ and outgoing 
lead is coupled to site $4$ of the molecular ring.
The transmission-energy characteristics under different cases of two input 
signals are presented in Fig.~\ref{fig2}(b), where the red, green and blue
lines correspond to the identical meaning as prescribed in Fig.~\ref{fig1}(b).
In a wide range of energy finite transmission is obtained when any one or both
inputs are ON, and here we set a specific energy $E=1.1\,$eV to reveal OR 
operation. The results are summarized in Table 2 under different cases of the
input signals which clearly describe the OR gate response.

\vskip 0.1cm
\noindent
{\bf \underline{NOT Gate}:} The molecular setup for designing NOT gate is
illustrated in Fig.~\ref{fig3}(a), where the benzene molecule is connected
to two leads (viz, lead-1 and lead-2) in {\em para configuration}. A gate
electrode is placed in the vicinity of site $3$ which serves as input of
the NOT gate and the output response is measured in lead-2. When the input
signal is OFF i.e., no gate voltage is applied in the gate electrode,
the transmission probability is finite (high) (red line) for the energy 
window shown in Fig.~\ref{fig3}(b). Whereas a sharp dip across $E=1.28\,$eV
in transmission curve (blue line) is noticed when the input signal is ON 
(viz, $\epsilon_3=1\,$eV). This is solely associated with the quantum 
interference effect of electronic waves passing through different arms of 
the molecular ring. Thus, setting the injecting electron energy at 
$E=1.28\,$eV the NOT gate operation is clearly visible, and the results
are summarized in Table 3.

To conclude, in the present work we address basic three logic operations
(OR, AND and NOT) considering a simple molecular structure. The main idea 
is that the molecular system having a loop geometry exhibits anti-resonant
states under a specific lead-to-molecule configuration, and these states 
are utilized in designing AND and NOT gates. Whereas for OR operation no
such anti-resonant states are required and thus easy to operate. We strongly
believe that the propositions can be implemented through an experimental
setup, and might be significant in the area of present molecular 
nanotechnology.

MP is grateful to University Grants Commission, India (F. 2-10/2012(SA-I)) 
for research fellowship.

\end{document}